\documentclass[twocolumn]{aastex6}
\usepackage{tikz,amsmath}
\usepackage{color}
\begin{document}
\newcommand{\Msol}{M$_{\odot}$}
\newcommand{\fxi}{f$_{esc}\xi_{int}$}
\newcommand{\gsim}{$\gtrsim$}
\newcommand{\lya}{Ly$\alpha$}
\newcommand{\kms}{km $s^{-1}$}
\newcommand{\ott}{$O_{32}$}
\newcommand{\hilight}[1]{\colorbox{yellow}{#1}}
\newcommand{\lsim}{$\lesssim$}
\newcommand{\rsrang}{$z\simeq2.3$}
\newcommand{\fesc}{$f_{esc}$}
\newcommand{\dittotikz}{%
    \tikz{
        \draw [line width=0.12ex] (-0.2ex,0) -- +(0,0.8ex)
            (0.2ex,0) -- +(0,0.8ex);
        \draw [line width=0.08ex] (-0.6ex,0.4ex) -- +(-1.5em,0)
            (0.6ex,0.4ex) -- +(1.5em,0);
    }%
}
\newcommand{\twocite}[2]{(\citealt{#1} cf.,~\citealt{#2})}
%%%%%%%%%%%% TITLE %%%%%%%%%%%%%%%%%%%%%%%%%%%%%%%%%%%%%%%%%%%%%%%%
%\title{\bf Lyman Continuum escape fraction of emission Line-selected, $\MakeLowercase{z}$\sim2.5 galaxies is less than $<20\%$.}
\title{\bf The Lyman Continuum escape fraction of emission line-selected $\MakeLowercase{z}\sim2.5$ galaxies is less than 15\%.}

\author{Michael J. Rutkowski\altaffilmark{1}; Claudia Scarlata\altaffilmark{2}; Alaina Henry\altaffilmark{3}; Matthew Hayes\altaffilmark{1}; Vihang Mehta\altaffilmark{2}, Nimish Hathi\altaffilmark{3,4}, Seth Cohen\altaffilmark{5}, Rogier Windhorst\altaffilmark{5}, Anton M. Koekemoer\altaffilmark{3}, Harry I. Teplitz\altaffilmark{6},  Francesco Haardt\altaffilmark{7,8}, Brian Siana\altaffilmark{9}}
\altaffiltext{1}{Department of Astronomy, AlbaNova University Centre, Stockholm University, SE-10691, SE}
\altaffiltext{2}{Minnesota Institute for Astrophysics, University of Minnesota, 116 Church St. SE, Minneapolis, MN 55455, USA}
\altaffiltext{3}{Space Telescope Science Institute, Baltimore, MD 21218, USA}
\altaffiltext{4}{Aix Marseille Universit\'{e}, CNRS, LAM (Laboratoire d'Astrophysique de Marseille) UMR 7326, 13388, Marseille, France} 
\altaffiltext{5}{School of Earth and Space Exploration, Arizona State University, Tempe AZ 85281, USA}
\altaffiltext{6}{Infrared Processing and Analysis Center, California Institute of Technology, Pasadena, CA 91125, USA}
\altaffiltext{7}{DiSAT, Universit\`{a} dell’Insubria, via Valleggio 11, 22100 Como, Italy}
\altaffiltext{8}{INFN, Sezione di Milano-Bicocca, Piazza delle Scienze 3, 20123 Milano, Italy}
\altaffiltext{9}{Department of Physics, University of California, Riverside, CA 92521, USA}
\date{\today}

%%%%%%%%%%%% ABSTRACT %%%%%%%%%%%%%%%%%%%%%%%%%%%%%%%%%%%%%%%%%%%%%%%%
\begin{abstract}

 Recent work suggests that strong emission line, star-forming galaxies
 may be significant Lyman Continuum leakers. We combine archival HST
 broadband ultraviolet and optical imaging (F275W and F606W,
 respectively) with emission line catalogs derived from WFC3 IR G141
 grism spectroscopy to search for escaping Lyman Continuum (LyC)
 emission from homogeneously selected $z\sim2.5$ SFGs. We detect no
 escaping Lyman Continuum from SFGs selected on [OII] nebular emission
 (N=208) and, within a narrow redshift range, on [OIII]/[OII]. We
 measure 1$\sigma$ upper limits to the LyC escape fraction relative to
 the non-ionizing UV continuum from [OII] emitters, \fesc\lsim5.6\%, and
 strong [OIII]/[OII]$>5$ ELGs, \fesc\lsim14.0\%. Our observations are
 not deep enough to detect $f_{esc}\sim 10$\% typical of the low
 redshift Lyman continuum emitters. However, we find that this
 population represents a small fraction of the star—forming galaxy
 population at $z\sim 2$. Thus, unless the number of extreme emission
 line galaxies grows substantially to $z\gtrsim 6$, such galaxies may be
 insufficient for reionization. Deeper survey data in the rest-frame
 ionizing UV will be necessary to determine whether strong line ratios
 could be useful for pre-selecting LyC leakers at high redshift.

\end{abstract} \section{Introduction}\label{sec:intro}  

%Star-forming galaxies (SFGs), quasars and active galactic nuclei (AGN)
%are the most probable sources of the far--UV background of Lyman
%continuum photons (LC, $\lambda<912\mbox{\AA}$) since in $z>10$ which
%have reionized neutral hydrogen in the inter-galactic medium (IGM) since
%$z>10$ \citep[seethe review][]{Loeb01}. 

Star-forming galaxies (SFGs) likely reionize neutral Hydrogen in the
early universe \citep[see review in][]{Loeb01}, when quasars are not
sufficiently numerous to contribute significantly to the ionizing
background \twocite{Ricci16}{Giallongo15}. Verifying this assumption by
directly measuring the ionizing output of Lyman Continuum
(LyC; $\lambda<912$\AA) is impossible---the IGM effectively attenuates all
LyC flux emitted along the line of sight to $z>6$ redshift galaxies.
Instead, the ionizing output of high redshift SFGs must be constrained
by surveys of low-redshift {\it analogs}, or indirectly
\citep[e.g.,][]{Jones13}. Criteria for pre-selecting LyC emitting
candidates from amongst the class of all SFGs that produce LyC are
crucial for such studies --- large, blind, deep surveys are infeasible
with the {\it HST}, the only telescope currently capable of obtaining
high resolution LyC imaging and spectroscopy.

Previously, pre-selection was made on actively SF galaxies \citep[young,
massive stars emit copious ionizing radiation, $Q_{H}
>$$10^{47}$s$^{-1}$;]{Schaerer03}. Studies with the HUT
\citep{Leitherer95} and the HST SBC \citep[e.g.,][]{Malkan03,Siana10} do
not detect escaping LyC. Large archival studies of LyC
emission from SFGs \citep{CBT09} and H$\alpha$-selected emission line
galaxies \citep{Rutkowski16} have generally reported non-detections,
likely indicating that the strong star-formation may be {\it conducive}
to LyC escape, but does not guarantee it. Until recently, few LyC
leakers were known; local starbursts Tol 0440-381, Tol 1247-232, Mrk 54,
and Haro11 \citep[e.g.]{Leitherer16,Puschnig16} and at $z$\gsim2, fewer
than $\sim$10, UV-selected star-forming galaxies
\citep[e.g.,][]{Mostardi16,Smith16,deBarros15}.

Recently, five (of five galaxies targeted) $z$\lsim0.3 compact
($r_e$\lsim 1 kpc) SFGs selected for their anomalously high nebular
oxygen ratios (\ott$\equiv$[OIII{\small$\lambda$}5007\AA] /
[OII{\small$\lambda\lambda$}3727,3729\AA]$>$5) have been confirmed as
LyC leakers \citep[\fesc$\sim$5-15\%;][]{Izotov16a,Izotov16b}, with an
additional 2-3 compact \ott\,galaxies predicted to be LyC leakers based
upon their Ly$\alpha$ profiles \citep{Verhamme17}. Furthermore,
\cite{Naidu16} applied a F275W-F336W color selection to identify 3 SFG
LyC leakers at $z\lesssim2$ in the HDUV (PID$:$13779; PI$:$P.Oesch),
each with \ott\gsim3. The success rate of LyC detection in \ott-emitters
makes this nebular-line diagnostic appealing for pre-selection. Here, we
investigate that potential utility.  In Section 2, we discuss the
selection of ELGs using HST IR grism spectroscopic catalogs combined
with rest-frame UV-optical imaging in the CANDELS fields. In Section 3,
we present new measurements to the absolute escape fraction, \fesc\,,for
these ELGs. We assume $\Lambda$CDM cosmology with $\Omega_m=0.27$,
$\Omega_{\Lambda}$=0.73, and H$_{0}=70 $km s$^{-1}$ Mpc$^{-1}$
\citep{Komatsu11}.\\ \vspace*{-20pt}

\section{Emission Line Galaxy Selection in CANDELS}\label{ss:lowzdata}

The HST WFC3/IR grism has been remarkably successful for surveying SFGs
with strong emission lines at $z\sim1-3$, as the
[OII]$\lambda\lambda$3726,3729\AA\, doublet, a well-calibrated signature
of star-formation \citep{Kewley04}, can be detected with the G141 grism
($\lambda_c\simeq1.4\mu$m) in 2\lsim$z$\lsim3.5 SFGs. The G141 can not
resolve this doublet, thus to avoid potential line mis-identification of
a candidate [OII], a photometric redshift is critical.   In the HST
CANDELS fields, the broadband UV-optical SED for galaxies is
well-sampled ensuring the robust identification of $z$\gsim2
[OII]-emitters. Unfortunately, rest-frame, broadband LyC imaging is not
available across the entire survey footprint. Thus, in our search here
for LyC emission from $z\sim2.5$ SFGs, we are limited to probing
$\sim$40\% of the area in GOODS-South and North. 

Within these regions, we select [OII]-emitters identified by the 3DHST
G141 grism survey. We selected ELGs requiring 1) SNR$_{\mbox{\small
OII}}>$3 and 2) within the redshift range $2.38<z<2.9$, where
$z\equiv$``{\tt z\_best}'' measured by \cite{Momcheva15} using both
grism and broadband photometry.  The lower redshift limit of this sample
is fixed to ensure the broadband (WFC3/UVIS F275W) imaging is strictly
sensitive to rest-frame LyC emission. In total, we select 208
[OII]-emitters (74, 109, and 25 in ERS, GOODS-N, and UVUDF,
respectively), with a mean (median) SFR=8.0 (3.9)\Msol$yr^{-1}$ and
stellar mass, M$_{\star}\simeq$10$^{9.9}$(10$^{9.6}$), respectively.
Included in this sample are 13 ELGs which were also considered in the
unpublished LyC survey in the ERS field presented by \citep{Smith16}.\\ \vspace*{-10pt}
\begin{figure}[htb]
 \centering
  \includegraphics[trim={0cm 6cm 0cm 6cm},clip,scale=0.42]{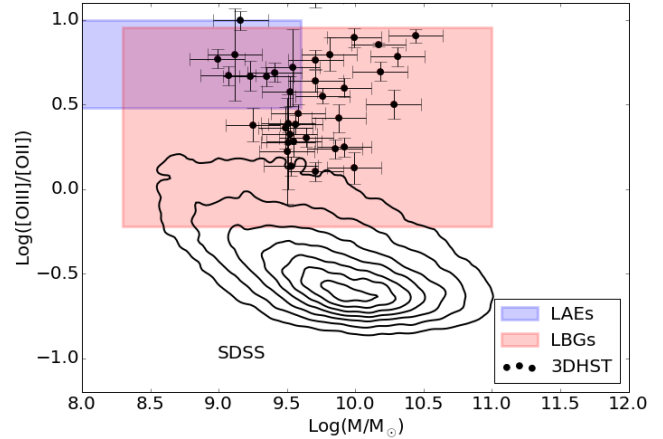} \\ \vspace*{-15pt}
\caption{We identify 41 $z\sim2.3$ \ott-emitters in the CANDELS fields.
Here, we plot \ott\,for the sources as measured from their G141 grism
spectra against stellar masses derived from SED fits in
\cite{Momcheva15}. We overplot contours indicating the
\ott\,distribution measured for SDSS galaxies \citep{Tremonti04}, and indicate with shaded regions the
parameter space populated by high redshift Ly$\alpha$ Emitters and Lyman Break Galaxies at
$z\sim2-3$ \citep{Nakajima14}.}
 \label{fig:o32plot} 
\end{figure} 

We note that within a narrow redshift range the G141 grism is
simultaneously sensitive to [OII] {\it and} [OIII]. Thus, we select a
second, independent sample of SFGs requiring$:$ 1) 2.25$<z<$2.31; 2)
SNR$_{[\mbox{\small OIII}]}>$3; and 3) SNR$_{\mbox{\small H}\beta}>$1.5.

This redshift range implies that the sample's LyC photometry could
include a contribution from non-ionizing emission in the bandpass. As
illustrated in Figure 2, the F275W throughput, T, \gsim1\% at
$\lambda<3086$\AA (910\AA at $z=2.39$). Only in the case of {\it zero}
attenuation by intervening neutral gas and dust (i.e.,
\fesc$\equiv$100\%) will the contribution redward of Lyman edge to the
F275W photometry be negligible ($<0.5\%$). The contribution by
non-ionizing photons to the measured ionizing flux will introduce a
systematic uncertainty to \fesc\,strictly less than unity measured using
the broadband method we apply in Section 3. Any candidate LyC leakers
identified in this sample must be considered tentative pending
spectroscopic followup. 
 
We identify 41 \ott\ emitters (22 and 19 in the GOODS-N and ERS
respectively). For the measurement of \ott, we require
[OIII]$\lambda$5007, which we derive from
[OIII]$\lambda\lambda$4959,5007\AA reported in 3DHST catalogs, applying
a uniform correction that assumes an intrinsic ratio of
$\lambda5007$/$\lambda4959$=2.98 \citep{Storey00} to correct for the
contribution of [OIII]$\lambda$4959. Of these \ott\,emitters, 13 are
identified with \ott$>5$. By comparison with the [OII]-emitter sample,
these ELGs have similarly high mean (median) SFR=6.8 (4.1)
\Msol$yr^{-1}$ and moderate stellar mass,
M$_{\star}\simeq$10$^{10.0}$(10$^{9.7}$).

In the following analysis of both samples, we use publicly-available
F275W imaging mosaiced by the individual survey teams (see Table
\ref{tab:thedata}). In GOODS-North, the HDUV team has prepared public
mosaics "v0.5" combining 5 of 8 HDUV pointings with CANDELS-Deep data. 
We note that the WFC3/UVIS is susceptible to significant ($\sim$50\%
losses) charge transfer inefficiencies. To mitigate this, UVUDF and
GOODS-North imaging programs (referenced in Col. 4 of Table
\ref{tab:thedata}) included a $\sim$10$e^{-}$ post-flash to minimize
charge losses. In the case of the ERS, these data were amongst the first
data obtained with the then newly-installed WFC3, and minimally affected
by the CTE \citep[see][for details]{Smith16}. In preparation for
analysis, we extracted 12\farcs$\times$12\farcs\ postage stamps from the
science and associated rms maps centered on each ELG. For uniformity,
all stamps were rebinned to a common pixel frame of 0\farcs09
pix$^{-1}$, the coarsest scale for which mosaics are available.

\begingroup
\tiny
\begin{longtable*}{lccccccc}
\caption{Archival F275W imaging}\\ 
\multicolumn{1}{c}{Field} &
\multicolumn{1}{c}{Survey} &
\multicolumn{1}{c}{Area$^a$} &
\multicolumn{1}{c}{Survey Depth (3$\sigma$)$^b$}  & 
\multicolumn{1}{c}{Reference}  &
\multicolumn{1}{c}{N$_{OII}$}  &
\multicolumn{1}{c}{N$_{\tiny O_{32}}$}  &
\multicolumn{1}{c}{N$_{\tiny  O_{32}>5}$}  \\ \hline \hline
\endhead
\multicolumn{8}{l}{{\sc Notes:} {\small $^a$--Approximate area in sq. arcminutes; $^b$--We report published point-source completeness limits [AB mag]}}\\
\multicolumn{8}{l}{but estimate the depth from the sky variance for the HDUV GOODS-N public mosaic.}\\
\multicolumn{8}{l}{{\small $^{c}$--For HDUV, mosaics at http://www.astro.yale.edu/hduv/DATA/v0.5/}}\\
\multicolumn{8}{l}{{\small \ddag--HST Program 13872 (Oesch et al.)}}\\
%\multicolumn{5}{l}{\,\,\,\,\,\,\,\,\,\,\,\,\,\,\,\,\,\,\,\,\,{\small of low-redshift interlopers.}}
\endlastfoot 
\hline \hline
GOODS-North & CANDELS-Deep & 120 & 27.8 & \cite{Koekemoer11} & 52 & 7 & 4\\ 
\hspace*{13pt}\dittotikz & HDUV & 70 & 27.9 & -\ddag- & 57 & 15 & 4\\ 
GOODS-South & UVUDF & 4.5 & 28.2 & \cite{Rafelski15} & 25 & --- & --- \\
\hspace*{13pt}\dittotikz & ERS & 60 & 26.5 & \cite{Windhorst11} & 74 & 19 & 5 \\ \hline \hline
\label{tab:thedata}
\end{longtable*}
\endgroup 
\vspace*{-20pt}
\section{The LyC escape fraction of \rsrang\,ELGs} 

\begin{figure}
\begin{center}
\includegraphics[trim=2cm 12cm 2cm 2cm,clip,scale=0.45]{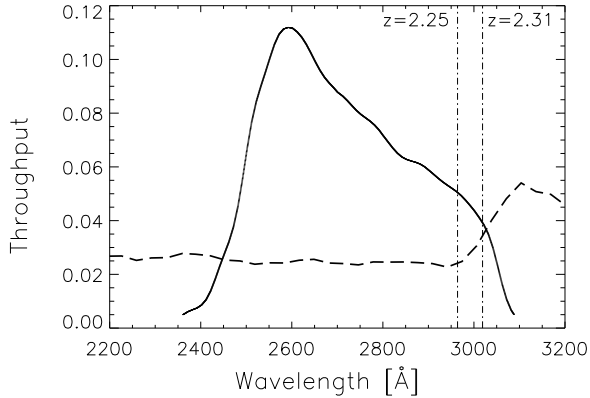} 
%produced on ramon in SANDBOX measure_integral.pro
\caption{The HST F275W broadband filter throughput, T, (solid curve)
declines to $<1$\% at $\lambda$\gsim3086\AA, corresponding to
$\lambda=910$\AA\,at $z>2.39$.  Thus, the Lyman edge falls within the
bandpass for SFGs selected at $2.25<z<2.31$ (indicated by dashed-dot vertical lines), implying the measurement of
LyC can be contaminated by non-ionizing emission. For a 1 Myr,
solar-metallicity burst simple stellar population model (with no
correction for attenuation by gas applied) from Bruzual \& Charlot
(2003) (dashed curve, scaled, in $f_{\nu}$), this contamination is
\lsim0.5\% {\it only} if \fesc=100\%. For \fesc$<$100\%, contamination
by the {\it non}-ionizing continuum to the LyC photometry will
necessarily increase to 100\% as \fesc\,decreases to zero.}
\label{fig:filtercurve} \end{center} \end{figure} 
%\citep{Cardamone09}. 

Broadband imaging surveys readily make {\it differential} measurements
of the ionizing (LyC) to non-ionizing (UV, measured at
$\lambda_{rest}\simeq1500$\AA) luminosity from galaxies.  At high
redshift, LyC from young stars within galaxies will be attenuated by
neutral gas and dust in the ISM, as well as by neutral HI in the IGM
along the line of sight.  This partly motivates a definition of the {\it
relative} escape fraction, following \cite{Steidel01}$:$

\begin{equation}
f_{esc,rel}=\frac{(L_{UV}/L_{LyC})_{int}}{(L_{UV}/L_{LyC})_{obs}}\cdot\exp[{\tau_{IGM}}],
\label{eqn:fescrel}
\end{equation}

$\tau_{IGM}$ is the (redshift-dependent) IGM attenuation of LyC by
neutral HI, typically modeled on measurements from absorption
line surveys towards high redshift bright quasars
\citep[e.g.,][]{Fardal98}. The intrinsic UV-to-LyC ratio must be modeled
for each galaxy individually, but typically ranges between 2-10 for
star-forming galaxies with ages less than $\sim10^8$yr. If the magnitude
of extinction due to dust in the ISM can be estimated from the SED, then
the {\it absolute} escape fraction can be directly related to
$f_{esc,rel}$ as$:$\vspace*{-10pt}

\begin{equation} 
f_{esc}=f_{esc,rel}\times\exp{[-\tau_{UV,dust}]}
\label{eqn:fesc} 
\end{equation}
\makeatletter
\setlength{\@fptop}{0pt}
\makeatother

We measured ionizing and non-ionizing photometry in the F275W and F606W
postage-stamps, respectively, using {\it Source Extractor} (Bertins \&
Arnouts 1996) in dual image mode, with the F606W as the detection
image\footnote{We use the relevant detection parameters
DETECT\_MINAREA=6, DETECT\_THRESH=3, BACK\_SIZE=10, BACK\_FILTERSIZE=5
and BACK\_FILTTHRESH=1.5, found by extensive testing to determine those
parameter that most accurately differentiated source from sky pixels in
the segmentation maps.}. 

The median F275W SNR for all ELGs is consistent with a statistical
non-detection ($\langle\,SNR\,\rangle=0.12$), as measured within each
ELG's corresponding F606W aperture defined in source extraction. A
visual inspection of {\it all} F275W stamps confirms that {\it no}
individual ELGs are LyC leakers, including those [OII]-emitters included
in the \cite{Smith16} sample. Note that for the median F606W
(rest-frame UV) continuum $m=$25 AB of this sample, the
surveys limits $f_{esc,rel}$\lsim2\% in the deepest (UVUDF) and
$f_{esc,rel}$\lsim12\% for the shallowest (ERS) mosaics. Smith et al.
report a detection of \fesc=0.14\% for the sample which overlaps with
the [OII]-emitter sample. Within the HDUV field no LyC
leakers have been previously identified\footnote{\cite{Naidu16}
identified 6 candidate LyC leakers, all at $z\simeq2$.}.

To measure \fesc, we apply the stacking procedure defined in
\cite{Siana10}, summing over all pixels in the F275W \& F606W stamps
associated with F606W-defined segmentation map. Furthermore, we sum in
quadrature the associated errors from the error maps, applying a (small)
correction for correlated noise introduced in the rebinning of the error
maps where necessary \citep{Casertano00}. This stacking yields no
statistically significant detections of LyC leakers. A visual inspection
of the associated stacked LyC frames (combined using IRAF {\tt
imcombine}; Figure 3) reveals no perceptible LyC flux within an aperture
defined by the non-ionizing UV image stack.   Here, we have cleaned all
LyC stamps before stacking, using the segmentation maps, to replace
pixels not associated with the ELG with randomly assigned pixel values
consistent with the sky background measured within each stamp. In Table
\ref{tab:fescvals}, we report \fesc\ for each stack as upper limits.

In rest-frame UV morphology, these galaxies are compact. For reference, an
aperture defined to include 90\% of the segmentation pixels common to
{\it all} galaxies has an area $\sim0.3$sq. arcsecond (physical radius,
r$\lesssim4$kpc), in good agreement with the measurements of $z\sim$2
galaxy sizes from \cite{Shibuya15}. Many galaxies ($>$70\%) do show
faint irregular UV features. Though we used the segmentation map
(defined by the rest-frame UV morphology) to define pixels to include in
the stacking of each galaxy, in principle these asymmetric low
surface brightness features could be lost to the sky when stacked.

\begin{figure} \begin{center} 
\includegraphics[trim={0cm 0cm 1cm 0cm},clip,scale=0.17]{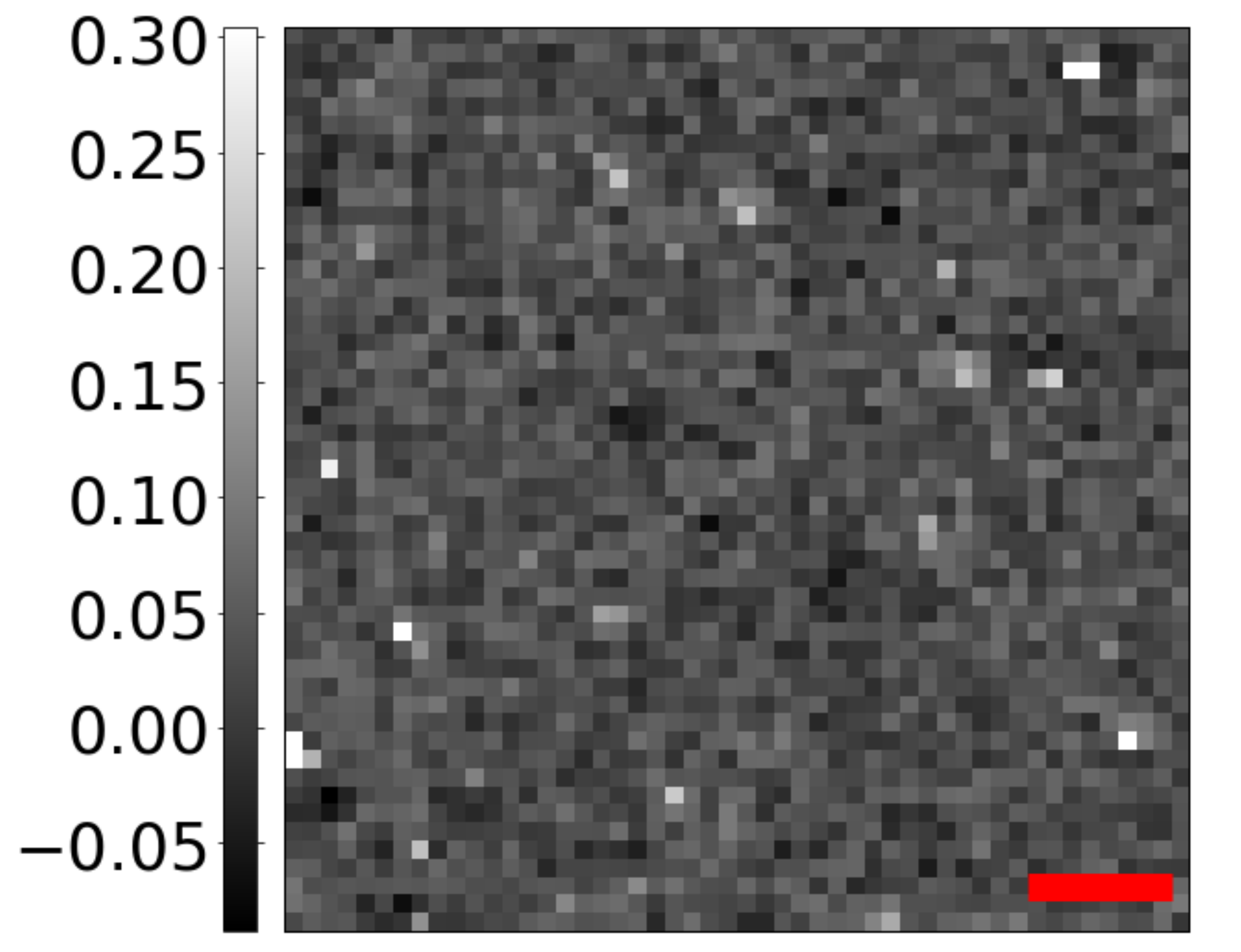}
\includegraphics[trim={1cm 0cm 0cm 0cm},clip,scale=0.17]{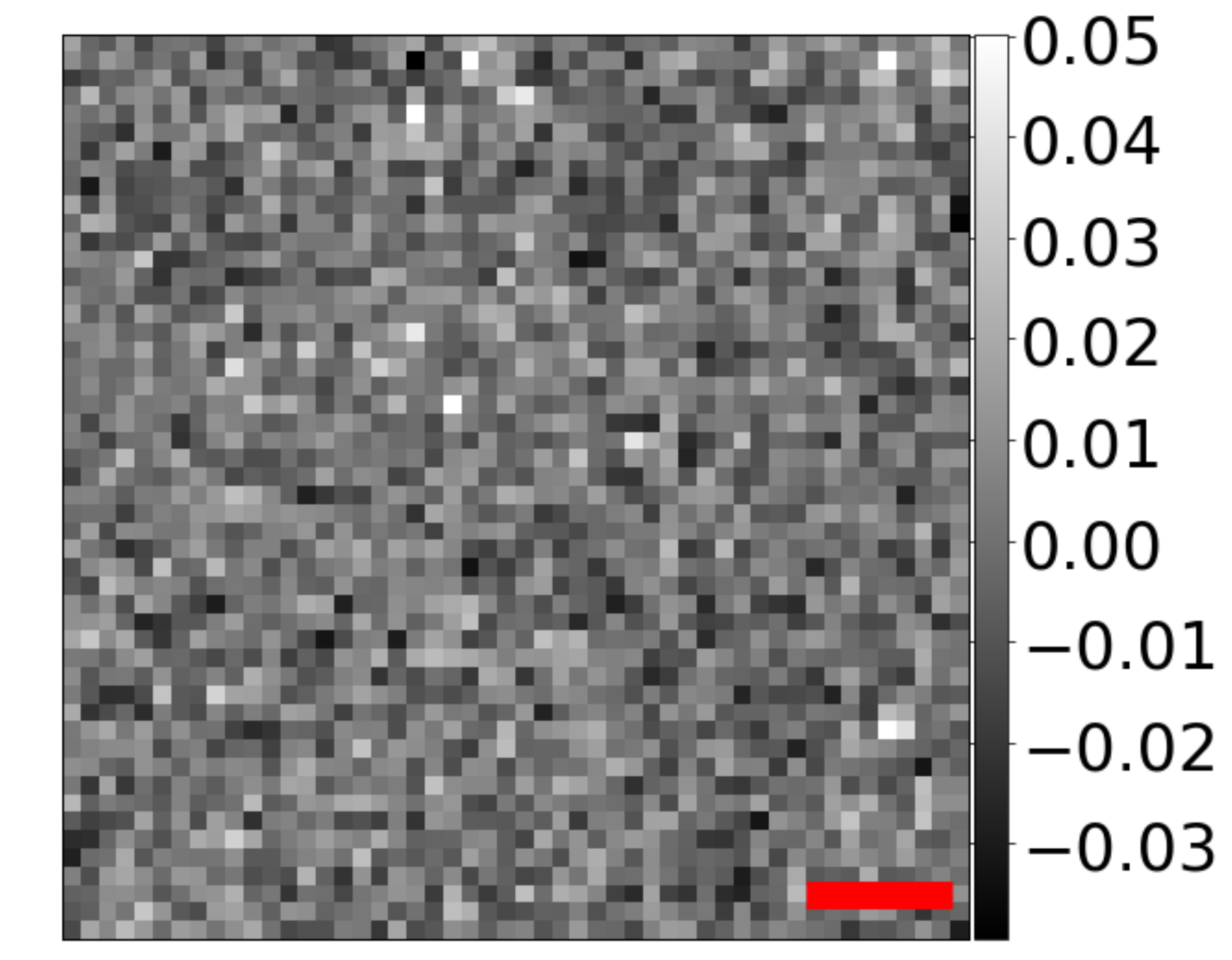} 
%%% fig3a is ~/Dropbox/PAPER_noAGN/SUBMITTED_VERSION] mjr% convert output_oii_sum_uh-huge.png fig3b.pdf
%%% fig3b is ~/Dropbox/PAPER_noAGN/SUBMITTED_VERSION] mjr% convert output_o32_sum_uh-huge.png fig3b.pdf

 \caption{The stacked rest-frame LyC images for the (a) [OII] and (b)
 strong (\ott$>$5) emitters.  A linear greyscale
 appropriately scaled for each stack is provided. A 1\farcs\ scalebar is
 overplotted (red) on each stacked image. All \fesc\ measurements are
 provided in Table \ref{tab:fescvals}.}
 \label{fig:allstacked}%\vspace*{-20pt} 
 \end{center}
  \end{figure}

We measure $f_{esc,rel}$, correcting each galaxy for IGM attenuation
using the correction factor from the piecewise parametrization of the
redshift distribution and column density of intergalactic absorbers
\citep[see][]{Haardt12}. For reference, exp[$\tau_{IGM}$]=1.72 (2.56), at
$z=$2.29 (2.56), the median redshift of the \ott\ ([OII])-selected
samples. Assuming $(L_{UV}/L_{LyC})_{int}=3$, appropriate for a young
($\sim$10$^7$yr), solar metallicity stellar population
\citep{Rutkowski16}, we measure $f_{esc,rel}$\lsim\,7.0, 7.8, and 18.9\%
(1$\sigma$) for the [OII]-, all \ott, and high \ott-selected samples. We
measure \fesc\, correcting for dust attenuation for each galaxy
individually. We measure $\tau_{UV, dust}$ assuming a \cite{Calzetti00}
reddening law ($R_V = 4.05$), and calculating stellar E(B-V) from the
best-fit $A_V$ measured from the broadband SED by \cite{Skelton14}.

We report \fesc$<$5.6\% for the [OII]-selected sample.  Note that the
upper limit on \fesc\ measured for random sub-samples of [OII]-emitters
drawn exclusively from the individual (unbinned) mosaics scale
approximately as $N^{-1/2}$, as expected from purely Poisson statistics.
Thus, in future work, using a re-reduction of all available F275W
imaging in the CANDELS fields to improve the size of the [OII]-selected
sample emitters, we will test for variations in \fesc\ in sub-samples
selected on, e.g., UV luminosity or inclination.

For the full \ott\-selected sample, we measure \fesc$<$6.7\%; for the
\ott$>$5, sample \fesc$<$14\%, consistent with the expectation for
Poissonian statistics if the extinction correction is appropriately
re-normalized to reflect the higher average extinction reported for the
\ott$>$5 sample. The mean IGM transmission for the \ott- and
[OII]-selected samples differs by a factor of $\sim$1.5, the
\ott\-emitters are intrinsically more luminous ($\sim3\times$), and the
possibility of a non-negligible contribution from non-ionizing flux in
the F275 bandpass (see Section 2) makes a direct comparison of \fesc\
upper limits for these samples more difficult. 

%\includegraphics[trim={1cm 0cm 9cm -1cm},clip,scale=0.20]{O32_OII_images.pdf} 
%l b r t 
\section{Discussion}\label{subsec:taues}

If these SFGs are analogs to the high redshift sources of reionization,
the measured upper limits can be informative. First, the 1$\sigma$ upper
limit to \fesc\ measured for [OII]-emitter sample is inconsistent with
the threshold of \fesc\gsim13\% required if high redshift SFGs reionize
the universe \citep[see][]{Robertson15} compatible with the independent
constraints on the ionization history of the IGM from the CMB (the
electron scattering opacity; $\tau_{es}$, see \cite{Planck16} and QSO
absorption line studies \citep{Mesinger07}. This tension is alleviated
considering the $3\sigma$ \fesc\, upper limit and noting that dwarf
galaxies less massive than these ELGs (with median
M$\simeq10^{9.5-10}$\Msol) are expected to contribute most significantly
to reionization \citep{Wise14,Robertson15}.  

Note \cite{Rutkowski16} measured, for (H$\alpha$-selected) $z\sim1$ SFGs,
\fesc$<4$\% (3$\sigma$). Selecting on more distant SFGs
using the same grism spectroscopy here, we are more sensitive to
intrinsically brighter line luminosities, $\sim\!4$ brighter
at $z\sim2.5$ than $z\sim1$, though intrinsically we can expect
[OII]/H$\alpha$\lsim1 \citep[$\simeq0.5$ at
$z\simeq0.1$;][]{Mouhcine05}.  As such, the average SFR for the
[OII]-selected sample is $\sim$2$\times$ that of the H$\alpha$ sample in
previous work, though the median SFR is measured for 3DHST sources from
the broadband SED in contrast to, e.g., \cite{Rutkowski16} which used
the extinction-corrected H$\alpha$ luminosity.  Thus, we caution any
strict interpretation of the \fesc\,upper limits derived here for
$z\simeq2.5$ SFGs and previous work at $z\simeq1$ as evidence for an
evolution in \fesc. \\ \vspace*{-10pt}

\begingroup
\tiny
\begin{longtable*}[t!]{lcccccccc}
\caption{Measured UV/LyC flux ratios$:$ Upper Limits to \fesc}\\
\multicolumn{1}{c}{Selection} &
\multicolumn{1}{c}{N$_{objs}$} &
\multicolumn{1}{c}{$\Delta(z)$} &
\multicolumn{1}{c}{Observed $f_{\nu,LyC}$$^a$} &  %NOT TRUE
\multicolumn{1}{c}{Observed $f_{\nu,UV}$} &
\multicolumn{1}{c}{IGM corr. UV/LyC} & 
\multicolumn{1}{c}{$f^{LyC}_{esc,rel}$$^b$} &
\multicolumn{1}{c}{$f^{LyC}_{esc}$} \\ \hline \hline
\endhead
\multicolumn{8}{l}{{\sc Notes:} {\small $^a$--Flux densities reported here in $\mu$Jy; $^b$--We assume $(L_{UV}/L_{LyC})_{int}$=3}; Italicized entries}\\
\multicolumn{8}{l}{\,\,\,\,\,\,\,\,\,\ {\small indicate non-detections and should be interpreted as limits.}}\\
%\multicolumn{5}{l}{\,\,\,\,\,\,\,\,\,\,\,\,\,\,\,\,\,\,\,\,\,{\small of low-redshift interlopers.}}
\endlastfoot 
\hline \hline
%\mbox{[OII]} & 208 & $2.38<z<2.9$ & 0.84$\pm$0.27 (3.1$\sigma$) & 9.12 & 41.49 & 21.0\% & 16.9\% \\ 
\mbox{[OII]} & 208 & $2.38<z<2.9$ & 0.45$\pm$0.27 (1.6$\sigma$) & 9.12 & $>$41.49 & $<$7.0\% & $<$5.6\% \\ 
All \ott & 41 & $2.25<z<2.31$ & 0.16$\pm$0.140 (1.1$\sigma$) & 4.53 & $>$38.44 & $<$7.8\% & $<$6.7\% \\  % significance=(1.14$\sigma$)
\ott$>5$ & 13 &  $2.25<z<2.31$ & -0.02$\pm$0.090 (-0.16$\sigma$) & 1.15 & $>$15.81 & $<$18.9\% & $<$14.0\% \\ \hline \hline
\label{tab:fescvals}
\end{longtable*} 
\endgroup 

Our observations are not deep enough to detect $f_{esc}\sim 10$\%
typical of the low redshift LyC emitters \citep{Izotov16b}, which have
comparably high nebular emission line ratios or similar star formation
rate surface densities, $\Sigma_{SFR}$\footnote{We measure
-2\,\lsim\,log($\Sigma_{SFR}$)\,\lsim\,1 [\Msol yr$^{-1}$ kpc$^{-2}$]
for \ott\-emitters, using the 3DHST broadband SFR and area from each
galaxy's F606W segmentation map.}. Our upper
limit on $f_{esc}$ derived for the high \ott\,ELGs is marginally
consistent ($f_{esc,rel} (3\sigma)\lesssim0.57$) with the detection of
LyC in a similar galaxy ({\it ion2}; $f_{esc,rel}=0.64^{1.1}_{-0.1}$)
studied by \cite{deBarros15}. 

We call attention to \fesc\, measured for the small number of SFGs
identified with high \ott\, emission, and the implication for the
contribution of their high redshift analogs to reionization. Generally,
reionization proceeds when a sufficient ionizing background can
be maintained by either a large number of relatively inefficient LyC
leakers or relatively fewer emitters which efficiently source LyC. The
number of ionizing background photons in a cosmological volume is
proportional to \fesc$\times\,n_{SFG}$, where $n_{SFG}$ is the volume
density of SF galaxies and \fesc$\simeq$10\% necessary for reionization.
However, not all SF galaxies are LyC leakers. In fact, it is well
established that the general population of SFGs have escape fraction
$<<10$\% \citep{Siana10,Grazian16}, and only the extreme \ott\,galaxies
appear to meet the requisite \fesc \citep{Izotov16b}. If $f_{leak}$ is
the fraction of SF galaxies that are LyC leakers, then the previous
relationship for the number of ionizing photons, N$_{ion}$, can be
rewritten as $N_{ion}\!\propto$\fesc$\times\,f_{leak}\times\,n_{SFG}$.
% the new calculations of the number O32 emitters in WISP is now at:OVI:/Users/mjrutkowski/HDUV/O32_SWING/NEWLINE 
In the WFC3 spectroscopic parallel survey\citep[WISP,][]{Atek10},
sensitive to both [OII] and [OIII] emission at $1.4$\lsim$z$\lsim$2.3$,
50\% of the catalogued galaxies are detected in both oxygen lines
\citep{Ross16}. Only $\sim$4\% of these sources are \ott$>$5 emitters.
With the upper limits presented here, assuming
$f_{leak}\sim4$\% and that this fraction does not evolve substantially
to $z\sim6$, such extreme objects would not support reionization.  High
redshift ($z>3$) SFGs do exhibit, on average, an enhanced ionization
state relative to low-redshift SFGs \citep[e.g.][]{Stanway14}, inferred
from the [OIII]/H$\beta$ ratio. Recently, \cite{Faisst16} modeled this
increased ionization state with redshift to predict the evolution of the
escape fraction evolution with redshift of \ott\,emitters, and found
such galaxies to be nearly sufficient to reionize the universe at
$z\sim6$.  Clearly, direct measurement of the median escape fraction for
strong emitters (\ott$>$5) with HST at $z<3$ is critical. This, in
combination with the direct measure of the evolution of the number
density of such extreme \ott\, galaxies towards the epoch of
reionization ($z>7$), a key result for JWST, will ultimately determine
whether such sources may reionize the universe.

\section{Conclusion}

We have combined archival high resolution HST UV imaging in the
rest-frame LyC for $z\sim2.5$ galaxies in the CANDELS deep fields,
selected on the presence of nebular oxygen emission lines in the 3DHST
IR grism spectra. We do not detect LyC escaping from [OII]- or
\ott-selected emitters individually. 

We stack the individual non-detections, and measure for each stack upper
limits to the absolute escape fraction less than 5.6, 6.7, and 14\%
(1$\sigma$), respectively.  Our limits on \fesc (3$\sigma$) for such
relatively massive galaxies do not rule out the possibility that SFGs
are able to sustain reionization. However, whether at $z\gtrsim 2$,
strong star formation and high \ott\ ratios alone are indicative of
significant LyC escape remains uncertain. Furthermore, we note that at
$z\sim2$ the class of galaxies with extreme \ott\,ratios remain
exceedingly rare.  In order for galaxies to be able to sustain
reionization, SFGs must evolve substantially from $z\sim 6$ to present,
such that at high redshift most have such highly ionized ISM conditions
indicated by the high \ott\,ratio. Such galaxies will be prime targets
for JWST at $z>3$ and future grism surveys and further constraints on
LyC emission from lower redshift \ott-selected ELGs will be important
for calibrating the evolution of LyC towards the epoch of reionization.
Deep HST surveys of large volumes at intermediate redshift will be
necessary to obtain the large sample sizes of strong \ott-emitters
necessary to determine whether LyC escape is linked to these observable
parameters such that their contribution can be meaningfully extrapolated
to the epoch of reionization probed by JWST. 

\acknowledgements 
This research was supported by NASA NNX13AI55G,
HST--AR Program \#12821.01 and GO-\#13352 using observations from
NASA/ESA HST, operated by the Association of Universities for Research
in Astronomy, Inc., under NASA contract NAS5-26555.  STScI is operated
by the AURA Inc., under NASA contract NAS5--26555.  M.H. acknowledges
the support of the Swedish Research Council (Vetenskapsr\r{a}det), the
Swedish National Space Board (SNSB), and the Knut and Alice Wallenberg
Foundation. RAW acknowledges JWST grants NAG5-12460 and NNX14AN10G from
NASA GSFC. This research has made use of the NASA ADS.

\clearpage


\begin{thebibliography}{99}
\bibitem[Atek et al.~(2010)]{Atek10} Atek, H., Malkan, M., McCarthy, P., et al., 2010, ApJ, 723, 104\vspace*{-2pt}
\bibitem[Bayliss et al.~(2014)]{Bayliss14} Bayliss, M. B.; Rigby, J., R., Sharon, K., et al., 2014, \apj, 790, 144\vspace*{-2pt}
\bibitem[Bertins \& Arnouts (1996)]{Bertins96} Bertin, E., \& Arnouts, S. 1996, A\&A, 117, 393\vspace*{-2pt}
\bibitem[Casertano et al.~(2000)]{Casertano00} Casertano, S., De Mello, D., Dickinson, M., 2001, AJ, 120, 2747\vspace*{-2pt}
\bibitem[Calzetti et al.~(2000)]{Calzetti00} Calzetti, D., Armus, L., Bohlin, R.C., et al., 2000, \apj, 533, 682\vspace*{-2pt}
\bibitem[Cardamone et al.~(2009)]{Cardamone09} Cardamone, C., Schawinski, K., Sarzi, M., et al., 2009, MNRAS, 399, 1191\vspace*{-2pt}
\bibitem[Cowie, Barger,\& Trouille (2009)]{CBT09} Cowie, L., Barger, A.J., \& Trouille, L., 2009, \apj, 692, 1476 \vspace*{-2pt}
\bibitem[de Barros et al.~(2015)]{deBarros15} de Barros, S., Vanzella, E., Amor\'{i}n, R., A\&A,585, A51\vspace*{-2pt}
\bibitem[Faisst (2016)]{Faisst16} Faisst, A., 2016, \apj, 829, 99\vspace*{-2pt}
\bibitem[Fardal, Giroux, \& Shull (1998)]{Fardal98} Fardal, M., Giroux, M.~L., Shull, J. M., 1998, AJ, 15, 2206\vspace*{-2pt}
\bibitem[Giallongo et al.~(2015)]{Giallongo15} Giallongo, E., Grazian, A., Fiore, F., et al.~2015, A\&A, 578, A83\vspace*{-2pt}
\bibitem[Grazian et al.~(2016)]{Grazian16} Grazian, A., Giallongo, E., Gerbasi, R., et al., 2016, A\&A, 585, A48\vspace*{-2pt}
\bibitem[Haardt \& Madau (2012)]{Haardt12} Haardt, F., \& Madau, P., 2012, \apj, 746, 125\vspace*{-2pt}
\bibitem[Inoue \& Iwata (2008)]{Inoue08} Inoue, A.K., \& Iwata, I., 2008, MNRAS, 387, 1681\vspace*{-2pt}
\bibitem[Izotov et al.~(2016a)]{Izotov16a} Izotov, Y.~I., Orlitov\'{a}, I., Schaerer, D., Nature, 529, 178\vspace*{-2pt}
\bibitem[Izotov et al.~(2016b)]{Izotov16b} Izotov, Y. I., Schaerer, D., Thuan, T. X., et al.  2016, MNRAS, 461, 3683\vspace*{-2pt} 
\bibitem[Jaskot \& Oey (2013)]{Jaskot13} Jaskot, A. \& Oey, M. S., 2013, \apj,766, 91\vspace*{-2pt} 
\bibitem[Jones et al.~(2013)]{Jones13} Jones, T.A., Ellis, R. S.,  Schenker, M.A., et al., 2013, \apj, 779, 52\vspace*{-2pt}
\bibitem[Kewley, Geller, \& Jansen~(2004)]{Kewley04} Kewley, L.Kewley, L. J., Geller, M. J., Jansen, R. A., AJ, 2004, AJ, 127, 2002\vspace{-2pt}
\bibitem[Koekemoer et al.~(2011)]{Koekemoer11} Koekemoer, A.M., Faber, S.M., Ferguson, H.C., et al., 2011, \apj, 197, 36\vspace*{-2pt}
\bibitem[Komatsu et al.~(2011)]{Komatsu11} Komatsu, E., Smith, K. M., Dunkley, J., et al.~2011, ApJS, 192, 18\vspace*{-2pt}
\bibitem[Leitherer et al.~(1995)]{Leitherer95} Leitherer, C., Ferguson, H.C., Heckman, T.M., Lowenthal, J.D., 1995, ApJL, 454, L19\vspace*{-2pt} 
\bibitem[Leitherer et al.~(2016)]{Leitherer16} Leitherer, C., Svea, Oey, , H.C., Heckman, T.M., Lowenthal, J.D., 1995, ApJL, 454, L19\vspace*{-2pt} 
\bibitem[Loeb \& Barkana (2001)]{Loeb01} Loeb, A., \& Barkana, R., 2001, ARA\&A, 39, 19\vspace*{-2pt}
\bibitem[Malkan, Webb, \& Konopacky (2003)]{Malkan03} Malkan, M., Webb, W., Konopacky, Q., 2003, \apj, 598, 878\vspace*{-2pt}
\bibitem[Mesinger \& Haiman (2007)]{Mesinger07} Mesinger, A., \& Haiman, Z., 2007, \apj, 660, 923\vspace*{-2pt}
\bibitem[Mouhcine et al.~(2005)]{Mouhcine05} Mouhcine, M., Lewis, I., Jones, B., et al., 2005, MNRAS, 362, 1143\vspace*{-2pt}
\bibitem[Momcheva et al.~(2015)]{Momcheva15} Momcheva, I., Brammer, G., van Dokkum, P., et al., 2015, arXiv$:$1510.02106\vspace*{-2pt}
\bibitem[Mostardi et al.~(2016)]{Mostardi16} Mostardi, R.E., Shapley, A.E., Steidel, C.C., et al.~2016, \apj, 810, 107\vspace*{-2pt}
\bibitem[Nakajima \& Ouchi (2014)]{Nakajima14} Nakajima, K. \& Ouchi, M., 2014, MNRAS, 442, 900\vspace*{-2pt}
\bibitem[Naidu et al.~(2017)]{Naidu16} Naidu, R. P., Oesch, P. A., Reddy, N., et al., 2017, subm. to \apj, arXiv eprint$:$1611.07038\vspace*{-2pt}
\bibitem[Planck Collaboration et al.~(2016)]{Planck16} Planck Collaboration, Adam, R., Aghanim, N., Ade, P.~A.~R., et al.~2015, arXiv$:$1605.03507\vspace*{-2pt}
\bibitem[Puschnig et al.~(2016)]{Puschnig16} Puschnig, J., Hayes, M., \"{O}stlin, G., et al., 2016, subm. to ApJ\vspace*{-2pt}
\bibitem[Ricci et al.~(2016)]{Ricci16} Ricci, F., Marches, S., Shankar, F., et al., 2016, MNRAS, arXiv$:$1610.01638\vspace*{-2pt}
\bibitem[Rafelski et al.~(2015)]{Rafelski15} Rafelski, M., Teplitz, H.I., Gardner, J.P., et al., 2015, ~AJ, 150,\#31\vspace*{-2pt}
\bibitem[Robertson et al.~(2015)]{Robertson15} Robertson, B. E.; Ellis, R. S.; Furlanetto, S. R., et al., 2015, \apj, 802, L19\vspace*{-2pt}
\bibitem[Ross et al.~(2016)]{Ross16} Ross, N.R., Malkan, M., Rafelski, M., et al., 2016, subm. to \apj\vspace*{-2pt}
\bibitem[Rutkowski et al.~(2016)]{Rutkowski16} Rutkowski, M.J., Scarlata, C., Haardt, F., et al., 2016, \apj, 819, 81\vspace*{-2pt}
\bibitem[Schaerer~(2003)]{Schaerer03} Schaerer, D., 2003, A\&A, 397, 527\vspace*{-2pt}	
\bibitem[Shibuya et al.~(2015)]{Shibuya15} Shibuya, T.m Ouchi, M., Harikane, Y., ApJS, 219,1\vspace*{-2pt}
\bibitem[Siana et al.~(2010)]{Siana10} Siana, B., Teplitz, H.I., Ferguson, H.C., et al.~2010, \apj, 723, 241\vspace*{-2pt}
\bibitem[Skelton et al.~(2014)]{Skelton14} Skelton, R., Whitaker, K.E., Momcheva, I.G., et al.,\, 2014, ApJS, 214, 24\vspace*{-2pt}
\bibitem[Smith et al.~(2016)]{Smith16} Smith, B., Windhorst, R.A., Jansen, R., et al. 2016, arXiv:1602.01555\vspace*{-2pt}
\bibitem[Stanway et al.~(2014)]{Stanway14} Stanway, E., Eldridge, J.J., Greis, S., 2014, MNRAS, 444, 3466\vspace{-2pt}
\bibitem[Steidel, Pettini, \& Adelberger~(2001)]{Steidel01} Steidel, C.C., Pettini, M., \& Adelberger, K.L., 2001, \apj, 546, 665\vspace*{-2pt}
\bibitem[Storey \& Zeippen~(2000)]{Storey00} Storey, P.~J. \& Zeippen, C.~J., 2000, MNRAS, 312, 813\vspace*{-2pt}
\bibitem[Tremonti et al.~(2004)]{Tremonti04} Tremonti, C., Heckmann, T., Kauffman, G., et al., 2004, \apj, 613, 898\vspace*{-2pt}
\bibitem[Verhamme et al.~(2017)]{Verhamme17} Verhamme, A., Orlitov\'{a}, I., Schaerer, D., et al., 2017, A\&A, 597, A13\vspace*{-2pt}
\bibitem[Windhorst et al.~(2011)]{Windhorst11} Windhorst, R.~A., Cohen, S.~H., Hathi, N.~P., et al., 2011, ApJS, 193, 27\vspace*{-2pt}
\bibitem[Wise et al.~(2014)]{Wise14} Wise, J.H., Demchenko, V. G., Halicek, M. T., et al., MNRAS,442, 2560\vspace*{-2pt}
%\bibitem[Zackrisson et al.~(2013)]{Zackrisson13} Zackrisson, E., Inoue, A.K., \& Jensen, H., 2013, \apj, 777, 39\vspace*{-2pt}
\end{thebibliography}
\end{document}